\documentclass[aps,prb,twocolumn,showpacs,superscriptaddress]{revtex4}
\usepackage{amssymb}
\usepackage{amsmath}
\usepackage{graphicx}

\begin{document}

\title{Greenberger-Horne-Zeilinger generation protocol for $N$
superconducting transmon qubits capacitively coupled to a quantum bus}
\author{Samuel Aldana}
\affiliation{Department of Physics, University of Basel,
Klingelbergstrasse 82, 4056 Basel, Switzerland}
\author{Ying-Dan Wang}
\affiliation{Department of Physics, University of Basel,
Klingelbergstrasse 82, 4056 Basel, Switzerland}
\affiliation{Department of Physics, McGill University, Montreal QC,
H3A 2T8, Canada}
\author{C. Bruder}
\affiliation{Department of Physics, University of Basel,
Klingelbergstrasse 82, 4056 Basel, Switzerland}

\begin{abstract}
We propose a circuit quantum electrodynamics (QED) realization of a
protocol to generate a Greenberger-Horne-Zeilinger (GHZ) state for $N$
superconducting transmon qubits homogeneously coupled to a
superconducting transmission line resonator in the dispersive limit. We
derive an effective Hamiltonian with pairwise qubit exchange
interactions of the $XY$ type, $\tilde{g}(XX+YY)$, that can be
globally controlled.  Starting from a separable initial state, these
interactions allow to generate a multi-qubit GHZ state within a time
$t_{\text{GHZ}}\sim \tilde{g}^{-1}$.  We discuss how to probe the
non-local nature and the genuine $N$-partite entanglement of the
generated state. Finally, we investigate the stability of the proposed
scheme to inhomogeneities in the physical parameters.
\end{abstract}



\maketitle

\section{Introduction}

Entangled quantum states are one of the essential resources for
quantum information processing. They are necessary for the
realization of quantum communication and the most important computational
tasks. Many efforts have been devoted to the elaboration of physical
systems enabling the generation and the control of such states. In
particular, different types of superconducting qubits are
promising candidates to solve this problem. Until recently limited to
two qubits
\cite{Steffen2006Sci,Plantenberg2007Nat,DiCarlo2009Nat,Ansmann2009Nat},
efforts to entangle superconducting qubits have lately reached a new
milestone with the experimental demonstration of three-qubit
entanglement \cite{DiCarlo2010Nat,Neeley2010Nat}.

In the present paper we consider transmon qubits
\cite{Koch2007PRA76,Schreier2008PRB77} in a circuit quantum
electrodynamics architecture
\cite{Blais2004PRA69,Wallraff2004Nat,Majer2007Nat, Schoelkopf2008Nat}
and present a way to generate GHZ states \cite{Greenberger1990AJP58},
i.e., maximally entangled states. Although the mathematical
description of multipartite entanglement for more than three qubits is
still debated
\cite{Verstraete2002PRA65,Lamata2007PRA75,Borsten2010PRL105}, GHZ
states remain paradigmatic entangled states which are, in particular,
useful for fault-tolerant quantum computing or quantum secret sharing
\cite{Hillery1999PRA59}.  So far, many different protocols have been
proposed to generate such states in circuit QED setups
\cite{Tsomokos2008NJP,Helmer2009PRA79,Hutchison2009CJP,Bishop2009NJP,Galiautdinov2009PRA80,Wang2010PRB81}.
Some of them are of probabilistic nature, i.e., if a measurement on
the $N$-qubit system has a specific result, the system is known to be
in a GHZ state after the measurement
\cite{Helmer2009PRA79,Hutchison2009CJP,Bishop2009NJP}.  In
Ref.~\onlinecite{Wang2010PRB81}, a M{\o}lmer-S{\o}rensen type
\cite{Molmer1999PRL82} one-step scheme to generate GHZ states both for
superconducting flux qubits and charge qubits was proposed.  The
procedure is independent of the initial state of the quantum bus and
works in the presence of multiple low-excitation modes. However,
higher excitation modes of the quantum bus will introduce
inhomogeneity because of the shorter wavelengths of the higher modes
and decrease the GHZ fidelity. Moreover, uncontrolled dissipation
might be coupled through the higher excitation modes and induce extra
noise. It would be ideal to devise a GHZ generation scheme that, while
keeping the one-step, deterministic nature, would involve only a
single mode of the quantum bus mediating the qubit interaction.

For this purpose,
in the present paper, we consider $N$ superconducting transmon
qubits homogeneously coupled to a superconducting transmission line
resonator in the dispersive limit, i.e., the architecture realized
in a number of experiments
\cite{Majer2007Nat,Filipp2009PRL102,DiCarlo2009Nat,DiCarlo2010Nat,Leek2010PRL104,Chow2010PRA81}.
We show that the system is characterized by effective qubit
exchange interactions of $XY$ type that can be globally controlled.
Starting from a separable initial state, these interactions allow to
generate a GHZ state in a deterministic one-step procedure.  We
discuss how to probe the non-local nature and the genuine $N$-partite
entanglement of the generated state and investigate the stability of
the proposed scheme to inhomogeneities in the physical parameters.
In contrast to Ref.~\onlinecite{Wang2010PRB81}, the qubit-resonator interaction
does not commute with the free Hamiltonian, and the qubit frequencies
are tuned close to one resonator mode. The time evolution of the
system is described by an effective Hamiltonian which
allows a direct implementation of the M{\o}lmer-S{\o}rensen idea.
Our scheme is the first one-step deterministic generation protocol of
GHZ states which could be possibly implemented in the currently
available circuit QED design.

The paper is organized as follows: in Section
\ref{fully_connected_network} we derive an effective Hamiltonian for
$N$ transmon qubits capacitively coupled to a superconducting
transmission line resonator in the dispersive regime. In Section
\ref{protocol} we describe the protocol for generating GHZ states in
our system. In Section \ref{measuring}, we discuss ways to confirm the
$N$-partite nature of the entanglement in the generated states, and in
Section \ref{inhomogeneous_coupling_frequencies}
we study the effects of non-ideal physical
parameters like inhomogeneities in the qubit-resonator coupling
constants.

\section{Fully connected network of transmon qubits in the dispersive limit}
\label{fully_connected_network}

We propose a solid-state implementation, based on an architecture of
superconducting transmon qubits capacitively coupled to a quantum bus
and derive an effective Hamiltonian for the system, which exhibits the
appropriate $XY$ exchange interaction.

Transmon qubits consist of a superconducting island
connected to a superconducting electrode through a Josephson tunnel
junction with capacitance $C_J$ and an extra shunting capacitance $C_B$.
 A gate voltage $V_g$ is applied to
the island via a gate capacitance $C_g$, allowing to tune the
dimensionless gate charge $n_g = C_g V_g/(2e)$. The system is
characterized by the charging energy $E_C= e^2/(2 C_{\Sigma})$,
where $C_{\Sigma}=C_g + C_J + C_B$ is the total capacitance of the island,
and the Josephson energy $E_J$ of the tunnel junction.

Such Josephson junction based qubits behave effectively as quantum
two-level systems in different regimes, categorized by the ratio
$E_J/E_C$. We will focus on the so-called transmon regime, when $E_J/E_C \sim 50-100$.
In this case the Hamiltonian of a single transmon qubit $\mathcal{H}_{\text{qb}}$
can be expressed as
\begin{equation}
\mathcal{H}_{\text{qb}} = 4 E_C(\hat n-n_g)^2 - E_J \cos \hat \varphi \,.
\end{equation}
In the following we assume that the Josephson junctions form a dc-SQUID
i.e., $E_J$ is tunable by an external applied magnetic flux
$\Phi_{\text{ext}}$ allowing to control independently each qubit. In
this case $C_{\Sigma} = C_g + 2 C_J^{(1)} + C_B$ and $E_J = 2 \tilde{E}_J
\cos(\pi \Phi_{\text{ext}}/\Phi_0)$ with $C_J^{(1)}$ and $\tilde E_J$
the capacitance and the Josephson energy of a single junction.

If a qubit is capacitively coupled to a superconducting transmission
line resonator, $C_g$ is now the capacitance between the
superconducting island and the resonator. In that particular situation
the gate voltage involves a dc-part and an extra term depending on the
state of the resonator, $V_g = V_g^{\text{dc}} + V(x)$. Therefore the
interaction with the resonator appears via the gate charge $n_g$,
which implicitly includes the voltage $V(x)$.
Transmon qubits are more robust to $1/f$-noise than charge qubits due
to their exponentially suppressed charge dispersion
\cite{Schreier2008PRB77}. However, we assume that the gate of each
qubit can be controlled separately by microwave pulses in order to
perform single qubit quantum gates.

For simplicity we consider the qubits to be coupled to a single mode
of the resonator. This is a reasonable assumption if the qubits are
nearly resonant with only one mode.  Since higher modes have
frequencies which are multiples of the fundamental frequency,
we can tune the qubit transition frequencies
such that the detuning with respect to one mode of the resonator is
one order of magnitude smaller than the detuning to all the other
modes. Under these conditions we can realize the dispersive limit for a
single mode of the resonator and neglect the influence of higher
modes, as is the case in experiments using one transmon
qubit~\cite{Bishop2009NPh}.

\begin{figure}
\centering
\includegraphics[width=0.98\linewidth]{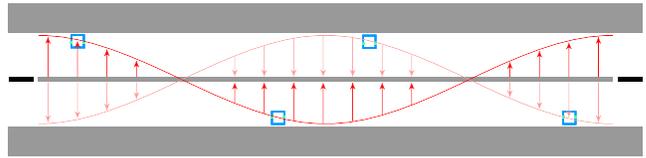}
\caption{(Color online) Sketch of a possible coplanar geometry for the
  proposed device with $N=4$ qubits. Qubits (small blue squares) are
  placed around the maxima of the electrical field amplitude (red line),
  i.e. near the center and the ends of the (quasi-)one-dimensional resonator
  (gray strip).  The second mode of the electrical field (red arrows)
  mediates the qubit-qubit interaction. Input and output ports
  of the resonator are drawn in black.}
\label{fig:schematic}
\end{figure}

For instance the qubits could be mainly coupled to the second mode
if they are placed near the ends or the center of the
resonator, that is the positions where the electrical field amplitude
is maximal. Such a possible geometry is sketched in Fig.~\ref{fig:schematic}.
Following the procedure of canonical quantization of a (quasi-) one-dimensional
superconducting resonator \cite{Blais2004PRA69}, the voltage across
the resonator is given by
\begin{equation}
V(x) = \sqrt{\frac{\omega_r}{L_0 c}} \cos \left(\frac{2\pi x}{L_0}\right)
(a + a^{\dag})\,.
\end{equation}
The length of the resonator is $L_0$ and its resonance frequency
$\omega_r = 2\pi/\sqrt{L_0^2 l c}$ depends on its capacity $c$ and
inductance $l$ per unit length. The position along the resonator is
denoted by $x\in [-L_0/2,L_0/2]$ and $a(a^{\dag})$ represent bosonic
annihilation (creation) field operators.

Following Ref.~\onlinecite{Koch2007PRA76}, the system, composed of the
resonator and $N$ transmon qubits, can be described with a generalized
Jaynes-Cummings Hamiltonian. This Hamiltonian is expressed in the
basis of transmon eigenstates $|j\rangle_q$, where the indices $q$
label the transmon qubits, and for readability we define the operators
$\Pi_j^{(q)}= |j\rangle _q \langle j|_q$,
$\sigma_{j,-}^{(q)}=|j\rangle_q \langle j+1|_q$,
$\sigma_{j,+}^{(q)}=|j+1\rangle_q \langle j|_q$ and set $\hbar=1$,
\begin{equation}
\mathcal{H} = \omega_r a^{\dag}a 
+\sum_{q=1}^N \sum_j \left[ \omega_j^{(q)} \Pi_j^{(q)}
+ \left( g_j^{(q)} \,a \, \sigma_{j,+}^{(q)} + \text{H.c} \right) \right].
\label{eq:JCham}
\end{equation}
The qubits frequencies $\omega_j^{(q)}$ are presumed to be tunable through external magnetic fields
$\Phi_{\text{ext}}^{(q)}$, changing the effective Josephson energies of the qubits
$E_J^{(q)}=2\tilde E_J^{(q)}\cos(\pi \Phi_{\text{ext}}^{(q)}/\Phi_0)$, and
the coupling frequencies $g_{j}^{(q)}$ depend on the position of the qubits.
Invoking the rotating-wave approximation, we have neglected rapidly oscillating terms.
In the transmon regime, we can only keep transmon-resonator coupling coefficients for
neighboring levels, since terms like $|i\rangle_q \langle j|_q$ for $|i-j|>1$ are comparatively small.
Moreover in the large $E_J/E_C$ limit asymptotic
expression can be obtained for $\omega_j^{(q)}$ and $g_{j}^{(q)}$ in first order
perturbation theory \cite{Koch2007PRA76},
\begin{equation}
\begin{gathered}
\omega_j^{(q)} \simeq \sqrt{8E_C^{(q)}E_J^{(q)}}\left(j+\frac{1}{2}\right)  -\frac{E_C^{(q)}}{12}(6j^2+6j+3),
\\
g_{j}^{(q)} \simeq g_0^{(q)} \sqrt{j+1}  \cos\left( \frac{2\pi x_q}{L_0}\right),
\\
 g_0^{(q)} \simeq -i \frac{e C_g^{(q)}}{C_{\Sigma}^{(q)}} \left( \frac{E_J^{(q)}}{2E_C^{(q)}} \right)^{1/4} \sqrt{\frac{\omega_r}{L_0 c}}.
\label{eq:frequencies}
\end{gathered}
\end{equation}
This form of the coupling frequencies $g_j^{(q)}$  describes the situation
shown in Fig.~\ref{fig:schematic}. The amplitudes of these coupling
coefficients $g_j^{(q)}$ can be assumed to be approximately homogeneous
if the positions $x_q$ of the qubits satisfy $|x_q/L_0| \simeq 0$ or $1/2$,
since the electrical field amplitude decreases quadratically with the distance
from its maxima and since the size of the qubits is typically much smaller
than the resonator wavelength in realistic systems. However even if close
to the center or the ends of the resonator, the qubits should be placed
sufficiently far apart to reduce direct inductive or capacitive qubit-qubit
coupling. There are also other positions that the qubits can be placed
in (e.g. nodes of higher modes). However, the homogeneity of the coupling
constants is important in our approach and should be taken care of.

In the so-called dispersive regime $|g_j^{(q)}/\Delta_j^{(q)}|\ll 1$, when transitions frequencies of the transmon qubits $\omega_{j,j+1}^{(q)}$
are detuned from the resonator frequency $\omega_r$, excitations of the resonator are virtual and the latter
will rather act as a quantum bus mediating effective qubit-qubit interactions.
The transition frequencies of the transmon qubits are defined as $\omega_{j,j+1}^{(q)}=\omega_{j+1}^{(q)}-\omega_{j}^{(q)}$
and their respective detuning as $\Delta_j^{(q)}=\omega_{j,j+1}^{(q)}-\omega_r$.
In this regime, eliminating the direct interaction between resonator and transmon qubits to lowest order in $g_j^{(q)}/\Delta_j^{(q)}$, 
we exhibit an effective qubit-qubit interaction. This can be seen by performing the canonical transformation $e^S\mathcal{H}e^{-S}$, where
\begin{equation}
S = \sum_{q=1}^N \sum_j \left( \frac{g_{j}^{(q)}}{\Delta_j^{(q)}} \; a
\; \sigma_{j,+}^{(q)}-\text{H.c}\right).
\end{equation}
Keeping terms up to second order in $g_j/\Delta_j$, we obtain.
\begin{equation}
\begin{aligned}
& e^S \mathcal{H} e^{-S}\\ 
\simeq &\; \left(\omega_r + \sum_{q=1}^N \left[ -\chi_0^{(q)} \Pi_0^{(q)} + \sum_{j\geq 1} \left(\chi_{j-1}^{(q)}-\chi_j^{(q)}\right)\Pi_j^{(q)}\right]\right)a^{\dag}a \\
& \;+ \sum_{q=1}^N \left[ \omega_0^{(q)} \Pi_0^{(q)} + \sum_{j\geq 1} \left(\omega_j^{(q)} + \chi_{j-1}^{(q)} \right) \Pi_j^{(q)} \right] \\
& \;+ \sum_{q=1}^N \left[ a a \sum_j \eta_j^{(q)} \sigma_{j+1,+}^{(q)}\sigma_{j,+}^{(q)} + \text{H.c} \right]\\
& \;+ \sum_{q\neq q'} \sum_{j,j'}\left[ \frac{\tilde g_{jj'}^{(qq')}}{2} \left( \sigma_{j,+}^{(q)} \sigma_{j',-}^{(q')} + \sigma_{j,-}^{(q)} \sigma_{j',+}^{(q')} \right) \right],
\end{aligned}
\label{eq:canonical}
\end{equation}
Here the ac-Stark shifts $\chi_j^{(q)}$, the two-photon transition rates $\eta_j^{(q)}$ and the effective qubit-qubit coupling coefficient $\tilde{g}_{jj'}^{(qq')}$ are given by
\begin{equation}
\begin{aligned}
\chi_j^{(q)} = &\; \frac{|g_j^{(q)}|^2}{\Delta_j^{(q)}},\\
\eta_j^{(q)} = &\; \frac{1}{2} \frac{g_j^{(q)}g_{j+1}^{(q)}}{\Delta_j^{(q)}\Delta_{j+1}^{(q)}}\left(\omega_{j,j+1}^{(q)}-\omega_{j+1,j+2}^{(q)}\right),\\
\tilde{g}_{jj'}^{(qq')} =  &\; \left|g_j^{(q)}g_{j'}^{*(q')}\right| \frac{\Delta_j^{(q)}+\Delta_{j'}^{(q')}}{2\Delta_j^{(q)}\Delta_{j'}^{(q')}}.
\end{aligned}
\end{equation}

Two-photon transitions can be safely neglected since the parameters
$\eta_j^{(q)}$ are small in the dispersive regime.  An effective
Hamiltonian $\mathcal{H}_{\text{eff}}$ is now obtained by restricting
our Hilbert space to the computational subspace, that is the first two
levels of each transmon qubit $\{|0\rangle,|1\rangle\}^{\otimes N}$.
In principle, the qubit-qubit interaction couples any states of the qubits
with more than one excitations to states that do not belong to the
computational subspace (e.g. for $N=3$, the state $|110\rangle$ or
$|111\rangle$ will be coupled to $|020\rangle$ or $|021\rangle$).
However, the amplitudes for these mixing processes of computational states
with such non-computational states are of order $g^2/(E_C\Delta)$ and will
be neglected~\cite{note_koch}. Under these conditions,

\begin{equation}
\begin{gathered}
\mathcal{H}_{\text{eff}} = \left(\omega + \sum_q \chi^{(q)} \sigma_z^{(q)}\right)a^{\dag}a + \sum_q \frac{\tilde\omega_{01}^{(q)}}{2}\sigma_z^{(q)} \\
+ \sum_{q,q'} \frac{\tilde{g}_{00}^{(qq')}}{4}\left( \sigma_x^{(q)}\sigma_x^{(q')}+\sigma_y^{(q)}\sigma_y^{(q')}\right),
\label{eq:effham0}
\end{gathered}
\end{equation}
where $\chi^{(q)} = \chi_0^{(q)}-\chi_1^{(q)}/2$,
$\sigma_z=\Pi_1-\Pi_0$, $\sigma_x=\sigma_+ + \sigma_-$,
$\sigma_y=-i(\sigma_+-\sigma_-)$. The resonator and qubit frequencies
get slightly renormalized, that is $\omega=\omega_r-\sum_q
\chi_1^{(q)}/2$ and
$\tilde\omega_{01}^{(q)}=\omega_{01}^{(q)}-\chi_0^{(q)}$. The
Hamiltonian has the desired YX-form, provided that all qubits have
identical parameters: that is all qubit and coupling frequencies are
homogeneous, $\tilde\omega_{01}^{(q)}=\Omega$, $|g_{0}^{(q)}|=g$,
$\Delta_0^{(q)}=\Delta$ and
$\tilde{g}_{00}^{(qq')}=\chi_0^{(q)}=\tilde g= g^2/\Delta$. Using
Eq.~(\ref{eq:frequencies}) we infer that $\chi^{(q)}=\chi=-\tilde g
E_C/(\Delta-E_C) < \tilde{g}$, where $E_C=\omega_{01}-\omega_{12}$ is
the weak anharmonicity of the transmon qubits. As mentioned earlier in
Eq.~(\ref{eq:frequencies}) the qubit transition frequencies can be
made homogeneous by tuning the flux biases $\Phi_{\text{ext}}^{(q)}$.
From now on we assume the $g_j^{(q)}$ are homogeneous. This is
motivated by a promising new transmon architecture
with tunable coupling that has been proposed recently
\cite{Srinivasan2011PRL106}. Inhomogeneous coupling constants will be
discussed in Sec. \ref{inhomogeneous_coupling_frequencies}.

Previous GHZ state generation protocols based on homodyne measurement
of the transmission
line~\cite{Helmer2009PRA79,Hutchison2009CJP,Bishop2009NJP} neglected
the effective exchange interaction because of the large differences in
qubit frequencies. In our case, the qubit frequencies $\omega_{01}^{(q)}$ are
tuned to be identical using the flux biases, and this effective
interaction plays a significant role in the generation of the GHZ
state in a one-step procedure as shown below.

If the qubit and coupling frequencies are homogeneous,
the total spin operators
$\hat J_{x,y,z}\!=\!\frac{1}{2}\sum_q \sigma_{x,y,z}^{(q)}$
and their corresponding Casimir operator ${\hat J}^2 = \hat J_x^2 +
\hat J_y^2 + \hat J_z^2$ can be used to write the effective
Hamiltonian in a very convenient form,
\begin{equation}
\mathcal{H}_{\text{eff}} = \omega a^{\dag}a + \tilde g \, \hat J^2 +
(\Omega + 2\chi a^{\dag}a) \hat J_z - \tilde g\,
\hat J_z^2.
\label{eq:effham}
\end{equation}
Evidently, $\mathcal{H}_{\text{eff}}$ is diagonal in the basis
$|J,J_z\rangle\otimes|n\rangle$, where the states $|J,J_z \rangle$,
describing the states of the $N$ qubits, are the eigenstates of the
operators $\hat J^2$ and $\hat J_z$ with respective eigenvalues
$J(J+1)$ and $J_z$. The states $|n\rangle$, describing the state of
the resonator, are eigenstates of $a^{\dag}a$ with eigenvalue $n\geq
0$. Since $[\mathcal{H},\hat J^2]=0$, any eigenstates of $\hat J^2$
will remain so under the action of this Hamiltonian. In the following,
we will restrict ourselves to such states with
$J\!=\!N/2$. For example states with all spins aligned
in a particular direction belong to this type and are therefore
an appropriate choice for the initial state. Setting $J\!=\!N/2$ in
what follows, we denote $|J\!=\!N/2,J_z\rangle$ by $|J_z\rangle$. The
eigenstates of $\mathcal{H}_{\text{eff}}$ are $|J_z\rangle \otimes
|n\rangle$ with eigenvalues $\varepsilon(n,J_z) = \omega n +
\tilde{g}(N/2+1)N/2 + ( \Omega + 2\chi n) J_z -
\tilde{g} J_z^2$.

\section{Protocol for generating GHZ states}
\label{protocol}
The effective Hamiltonian $\mathcal{H}_{\text{eff}}$ allows to produce
GHZ states by turning on the interaction for a definite duration
$t_{\text{GHZ}}$. It was shown in
Refs.~\onlinecite{Molmer1999PRL82,Galiautdinov2009PRA80} that a Hamiltonian of
the type $\tilde{g} \hat J_{x}^2$ will produce a GHZ state after the
time $\pi/(2\tilde{g})$, starting for instance from the state
$\bigotimes_q|1\rangle_q$. Implementation of such scheme in other qubit system
has also been proposed \cite{Helmerson2001PRL87,Zheng2001PRL87}.
The multi-qubit gate $\exp(i\pi\hat
 J_x^2/2)$ is sometimes referred to as M{\o}lmer-S{\o}rensen gate.

We conveniently choose an initial state with all the qubits
aligned in the same direction, that is the maximal superposition state $|\psi_0 \rangle =
\bigotimes_q \left(|0\rangle_q+|1\rangle_q\right)/\sqrt{2}$. We assume that the qubits and the resonator
are initially in a product state and the state of the resonator at
$t=0$ is denoted $\rho_{\text{res}}$,
\begin{equation}
\rho(t\!=\!0) = |\psi_0 \rangle \langle \psi_0|\otimes
\rho_{\text{res}}\,.
\end{equation}
Moreover $|\psi_0\rangle=|J_x\!=\!N/2 \rangle$ and can be expressed as a
linear superposition of the states $|J_z\rangle$
(see Appendix \ref{app:schwinger}),
\begin{equation}
|\psi_0 \rangle = \frac{1}{\sqrt{2^N}} \sum_{k=0}^N
\sqrt{\textstyle{\binom{N}{k}}} \; |J_z\!=\!k\!-\!N/2\rangle\,.
\label{eq:initialstate}
\end{equation}

We define $\rho(t)$ as the density matrix evolving under
the action of the time-evolution operator $U(t) =
\exp(-i\mathcal{H}_{\text{eff}} t)$, where $\mathcal{H}_{\text{eff}}$ is
the effective Hamiltonian Eq.~(\ref{eq:effham}),
\begin{equation}
\rho(t) = U(t) \,\rho(t\!=\!0)\, U^{\dag}(t).
\label{eq:transfrho}
\end{equation}
We assumed that $g/\Delta \ll 1$ and therefore we have neglected the effect
of the canonical transformation $e^S$ on the state vector.
This turns out to be particularly useful,
since $U(t)$ is diagonal in the basis $|n\rangle$, we can describe
directly the time evolution of the reduced density matrix of the
qubits $\rho_{\text{qb}}(t)$, obtained by tracing over the resonator
state,
\begin{equation}
\begin{aligned}
& \rho_{\text{qb}}(t) := \text{Tr}_{\text{res}} [\rho(t)] \\ &
  = \frac{1}{2^N} \sum_{n,k,k'} \langle n |\rho_{\text{res}}|n\rangle
  \sqrt{\textstyle{\binom{N}{k} \binom{N}{k'}}} e^{-i
    (\varphi_{k,n}(t) - \varphi_{k',n}(t))} \\ & \qquad \qquad \qquad
  |J_z\!=\!k\!-\!N/2\rangle \langle J_z\!=\!k'\!-\!N/2|\,,
\label{eq:rhot}
\end{aligned}
\end{equation}
where $\varphi_{k,n}(t) = k\left( \Omega t + 2\chi t n+\tilde{g}t(N-k)
\right)$.

The Greenberger-Horne-Zeilinger (GHZ) states we aim to produce are of the
following form,
\begin{equation}
|\text{GHZ}^{\pm}\rangle = \frac{1}{\sqrt{2}} \left( \bigotimes_{q=1}^N
\frac{|0\rangle_q+|1\rangle_q}{\sqrt{2}} \pm i \bigotimes_{q=1}^N
\frac{|0\rangle_q-|1\rangle_q}{\sqrt{2}} \right)\,,
\label{eq:defghz}
\end{equation}
which are standard GHZ states up to single qubit rotations. These states can be expressed as a linear
superposition of the states $|J_z\rangle$ as well (see Appendix \ref{app:schwinger}):
\begin{equation}
|\text{GHZ}^{\pm}\rangle = \sum_{k=0}^N \frac{1 \pm i\;e^{i \pi
    k}}{\sqrt{2^N}\sqrt{2}} \sqrt{\textstyle{\binom{N}{k}}}
|J_z\!=\!k\!-\!N/2\rangle\,.
\label{eq:ghzx}
\end{equation}
To see why a GHZ state is produced after some time $t_{\text{GHZ}}$ we
consider the effects of either $\exp(i\tilde g t \hat J_z^2)$ or
$\exp[i\tilde g t(\hat J_z^2-\hat J_z)]$ (for $N$ either even or odd)
on the state $|J_z\!=\!k\!-\!N/2 \rangle$. We establish that one of
the two possible GHZ states Eq.~(\ref{eq:defghz}) is produced when $\tilde
g t=\pi/2$ by noticing that
\begin{equation*}
\begin{gathered}
\frac{1 + i e^{i\pi (k+\frac{N}{2}-1)}}{\sqrt{2}} = e^{-i\frac{\pi}{4}
  + i \frac{\pi}{2}(k-\frac{N}{2})^2}\,,\quad N\text{ even}\,,
\\ \frac{1 + i e^{i\pi (k+\frac{N-1}{2})}}{\sqrt{2}} =
e^{-i\frac{\pi}{8}+i\frac{\pi}{2} \left[ (k-\frac{N}{2})^2 -
  (k-\frac{N}{2}) \right]}\,,\quad N\text{ odd}\,.
\end{gathered}
\end{equation*}
Thus, a GHZ state is produced for every odd multiple of time $t_{\text{GHZ}}$.
The shortest preparation time is $t_{\text{GHZ}} = \pi/(2\tilde g)$

However the remaining term of the effective Hamiltonian in
Eq.~(\ref{eq:effham}), the one which is proportional to $\hat{J}_z$,
induces a collective rotation of the final state. The rotation angle depends again on $N$ and
the state of the resonator. The state $\rho_{\text{qb}}(t_{\text{GHZ}})$ is,
\begin{equation}
\rho_{\text{qb}}(t_{\text{GHZ}}) = \sum_n \langle n|\rho_\text{res} |n\rangle
\;| \text{GHZ}(\alpha_n) \rangle \langle \text{GHZ}(\alpha_n)|\,.
\label{eq:rhotghz}
\end{equation}
Here,
\begin{equation}
\begin{aligned}
| \text{GHZ}(\alpha) \rangle
= e^{-i\alpha \hat J_z} \frac{1}{\sqrt{2}} & \left( \bigotimes_{q=1}^N
\frac{|0\rangle_q + |1\rangle_q}{\sqrt{2}} \right.
\\
& \left. + e^{i\pi\frac{N-1}{2}} \bigotimes_{q=1}^N
\frac{| 0\rangle_q - |1\rangle_q}{\sqrt{2}}
\right) ,
\label{eq:localrotations}
\end{aligned}
\end{equation}
and $2\alpha_n/\pi = (\Omega + 2n\chi) /\tilde{g} $ for $N$ even.  For
$N$ odd, $2\alpha_n/\pi = (\Omega + 2n\chi) /\tilde{g} -1 $, and the relative
phase $\exp(i\pi(N-1)/2)$ in Eq.~(\ref{eq:localrotations}) is
changed to $\exp(i\pi N/2)$.

We notice that the produced states $\rho(t_{\text{GHZ}})$ is not
exactly the state depicted in Eq.~(\ref{eq:defghz}) and therefore
certain constraints on the angles $\alpha_n$ in Eq.~(\ref{eq:rhotghz})
are required to generate the proper state
$|\text{GHZ}^+\rangle$.  At low temperature, only the ground state of
the resonator is significantly populated and $\langle 0|\rho_{\text{res}}|0\rangle
\gg \langle n|\rho_{\text{res}}|n\rangle$ for $n\geq1$. Thus we can
restrict our considerations to $\alpha_{n=0}$ and this translates to some
condition on the ratio $\Omega/\tilde{g}$.

To illustrate this we consider the resonator to be initially in its ground state
$\rho_{\text{res}} = |n\!=\!0\rangle\langle n\!=\!0|$. The state
$|\text{GHZ}^+\rangle$ is indeed produced at $t_{\text{GHZ}}$,
provided we can tune the frequencies $\Omega$ and $\tilde g$ such that
\begin{equation}
\frac{\Omega}{\tilde g} = 4m+2-N\,,\qquad m \in \mathbb{Z}\,.
 \label{eq:Omegaconditon2}
\end{equation}
If the above condition cannot be satisfied, some correcting pulse
$\exp(i\delta_N \hat J_z)$ can be applied to the final state
$\rho_{\text{qb}}(t_{\text{GHZ}})$ to obtain a proper $|\text{GHZ}^+ \rangle$
state. The appropriate pulse length $\delta_N$ depends on $N$ and the
ratio $\Omega/\tilde{g}$,
\begin{equation}
\delta_N = \frac{\pi}{2} \left[ \left( \frac{\Omega}{\tilde{g}} + N -2
  \right)\,\text{mod}\,4 \right] \,.
\label{eq:correctingpulse}
\end{equation}
Furthermore $\delta_N=0$ implies Eq.~(\ref{eq:Omegaconditon2}).

If not only the ground state of the resonator is populated, higher
photon numbers $n$ produce rotated GHZ states, according to
Eq.~(\ref{eq:rhotghz}). We notice that $\langle
\text{GHZ}(\alpha_n)|\text{GHZ}(\alpha_{n+k})\rangle = 
\cos^N(k\pi\chi/(2\tilde{g}))$, which means that if a $|\text{GHZ}^+
\rangle$ state is produced for excitation number $n$, a slightly
rotated state
$\exp(-i\pi\chi\hat J_z/\tilde{g})|\text{GHZ}^+\rangle$ is
produced for $n+1$ (since $\chi<\tilde{g}$). 
Assuming some correcting pulse $\exp(i\delta_N
\hat J_z)$ has been applied, the reduced density matrix of the qubits
$\rho_{\text{qb}}$ is a mixture of rotated GHZ states with classical
probabilities depending only on the initial state of the resonator,
\begin{equation}
\begin{aligned}
& e^{i \delta_N \hat J_z} \rho_{\text{qb}}(t_{\text{GHZ}}) e^{-i \delta_N \hat J_z}
\\
& = \langle 0 |\rho_{\text{res}}|0\rangle |\text{GHZ}^+\rangle \langle \text{GHZ}^+|
\\
&  + \sum_{n>0} \langle n |\rho_{\text{res}} |n\rangle e^{-i \pi n \frac{\chi}{\tilde{g}}\hat J_z} |\text{GHZ}^+\rangle \langle \text{GHZ}^+| e^{i \pi n\frac{\chi}{\tilde{g}}\hat J_z} \,.
\label{eq:correctedrho}
\end{aligned}
\end{equation}

We will now show that it is possible to choose realistic physical
parameters in agreement with our assumptions. Transmon
qubits have typical frequencies $\Omega/2\pi$ around 10 GHz and
coplanar waveguide resonators (the quantum bus) can be realized with
frequencies $\omega/2\pi$ of the order of 10 GHz with high quality
factors \cite{Goppl2008JAP}. Transmon-resonator coupling frequencies $g/2\pi$ around 200
MHz is a reasonable assumption. Detuning the qubits from the resonator such
that $g/\Delta \simeq 1/10$ would lead to an effective qubit-qubit coupling
of the order of $\tilde g = g/10$ and to preparation
time $t_{\text{GHZ}}$ of approximately 12.5 ns.

\section{Measuring the generated GHZ states}
\label{measuring}
The question of detecting and probing the states generated in our
scheme naturally arises. For $N \geq 4$, there is no unique way to
quantify entanglement. We will focus on a measurement of the
Bell-Mermin operator \cite{Mermin1990PRL65} defined as
\begin{equation}
\begin{aligned}
B = &\; \frac{e^{i\pi N}}{2 i} \left[ \bigotimes_{q=1}^N \left(\sigma_{z}^{(q)}-i\sigma_y^{(q)}\right) - \bigotimes_{q=1}^N
\left(\sigma_{z}^{(q)}+i\sigma_y^{(q)}\right) \right]
\\
= &\; 2^{N-1}\left( |\text{GHZ}^+\rangle
\langle\text{GHZ}^+| - |\text{GHZ}^-\rangle \langle\text{GHZ}^-|
\right)\,,
\end{aligned}
\label{eq:definitionbm}
\end{equation}
whose expectation value for $N$-qubit quantum states is bounded by
$|\langle B \rangle |\leq 2^{N-1}$, and the extremal values $\pm
2^{N-1}$ are reached for the states $|\text{GHZ}^{\pm} \rangle$. The
maximal value predicted by local hidden-variable theory
is $\sqrt{2^N}$ ($\sqrt{2^{N-1}}$) for $N$ even (odd), leading to an
exponentially increasing violation for the states $|\text{GHZ}^{\pm}
\rangle$ with $N$, the number of qubits. Therefore, a measurement of
the Bell-Mermin operator leading to a result greater than
$\sqrt{2^N}$ ($\sqrt{2^{N-1}}$) indicates the non-local nature of
the generated quantum states.

Other bounds can be derived for this operator: e.g., any separable
state $\rho^{S}$ satisfies $|\text{Tr}(\rho^SB)|\leq 1$.  A
significant bound can also be derived if the state is $m$-separable,
i.e.  describes a system which is partitioned in $m$ subsystems that
only share classical correlations. In other words, a {\it pure} state is
called $m$-separable, for $1<m\le N$, if it can be written as a product
of $m$ states,
\begin{equation}
|\psi^m \rangle = \bigotimes_{i=1}^m |\psi_i \rangle_{P_i},
\end{equation}
where the $\{P_i\}$ describe a partition of the $N$ qubits.  Thus, a
separable state in the traditional sense is $N$-separable.  A {\it
  mixed} $m$-separable state $\rho^m$ is defined as a convex sum of
pure $m$-separable states, which might belong to different partitions
\cite{Guehne2009PR}.  Such an $m$-separable state satisfies
$\text{Tr}(\rho^m B)\leq 2^{N-m}$.  Thus, any measurement of $B$ with
outcome above $2^{N-2}$ indicates that the state is not even
biseparable ({\it 2-separable}) and demonstrates the existence of
genuine $N$-partite entanglement.

The Bell-Mermin operator expectation value can in principle be
obtained experimentally.  This operator can be expressed as a sum of
parity operators, and inferring its expectation value would require
$2^{N-1}$ parity measurements,
\begin{equation}
\langle B \rangle = \sum^N_{l=1\;\text{odd}} \sum_{p}
(-1)^{N-\frac{l+1}{2}} \left\langle \bigotimes_{q=1}^{N-l} \sigma_{z}^{p(q)}
\bigotimes_{q'=N-l+1}^{N} \sigma_{y}^{p(q')} \right\rangle.
\label{eq:permutations}
\end{equation}
For each term, $l$ is the number of factors $\sigma_y$ and $\sum_p$
stands for the sum over the $\binom{N}{l}$ permutations $p$ that give
distinct products. The states $|\text{GHZ}^{\pm}\rangle$ defined in
Eq.~(\ref{eq:defghz}) are those that give exactly $\pm1$ for each of
the $2^{N-1}$ terms.

There are therefore $2^{N-1}$ parity measurements to realize which is
possible only if one is able to generate GHZ states with high accuracy
in a repeated way. Following Ref.~\onlinecite{Hutchison2009CJP}, these
parity operators could be measured by dispersive readout. Since the
frequency of the resonator is ac-Stark shifted $\omega \to \omega +
\chi \sum_q \sigma_{z}^{(q)}$, it is possible to access the value
of the operator $\hat J_z$. The value of the parity operator $\bigotimes_q
\sigma_{z}^{(q)}$ can then be unambiguously deduced from $J_z = \langle
\hat J_z \rangle$,
\begin{equation}
\left\langle \bigotimes_{q=1}^N \sigma_{z}^{(q)} \right\rangle =
(-1)^{\frac{N}{2}-J_z}\,.
\end{equation}
Hence, we can measure all the needed parities by rotating the
operators $\sigma_{y}^{(q)}$ appearing in
Eq.~(\ref{eq:permutations}) to $\sigma_{z}^{(q)}$ using
single-qubit rotations.

\begin{figure}
\centering
\includegraphics[width=\linewidth]{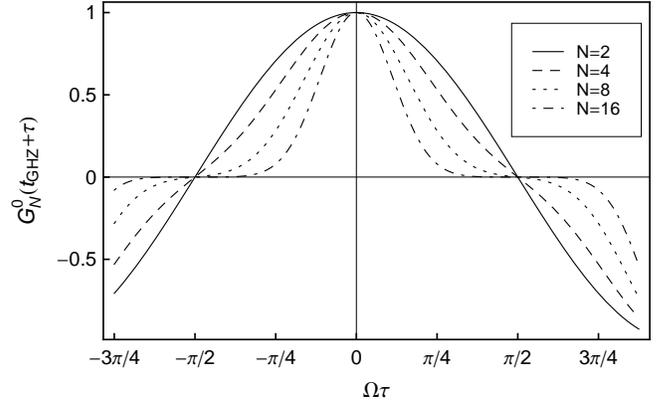}
\caption{Behavior of the function $G_{N}^0(t_{\text{GHZ}}+\tau)$
for different $N$, assuming for simplicity that $\delta_N=0$.}
\label{fig:GNn}
\end{figure}

By means of Eq.~(\ref{eq:rhot}), we can give an expression for the
time evolution of the expectation value of the Bell-Mermin operator, $\langle
B(t) \rangle = \text{Tr}\, \left[ B \rho_{\text{qb}}(t)
  \right]$. For this purpose we can express the matrix elements of
$B$ in the basis of the states $|J_z\rangle$, which
diagonalizes the effective Hamiltonian,
\begin{equation}
B = \sum_{k,k'=0}^N b_{k,k'}\; | J_z\!=\!k'\!-\!N/2\rangle
\langle J_z\!=\!k\!-\!N/2 | \,,
\label{eq:matelembm}
\end{equation}
where
\begin{equation}
 \\ b_{k,k'} = \frac{1}{2i}
\sqrt{\textstyle {\binom{N}{k}\binom{N}{k'}}} \left[ (-1)^k - (-1)^{k'}
\right]\,.
\end{equation}
Hence, $\langle B(t)\rangle$ can be expressed as a sum of oscillating
functions $G_{N}^n$, indexed by the photon number $n$,
\begin{equation}
\langle B(t) \rangle = 2^{N-1} \sum_{n=0}^{\infty} \langle n |
\rho_{\text{res}}|n\rangle \; G_{N}^n(t)\,.
\label{eq:BM_oscillations}
\end{equation}
The functions $G_{N}^n$ are Fourier series
over a finite range of frequencies $\tilde \omega_{k,k'}^n$ defined as
$\tilde \omega_{k,k'}^n= (k-k')\left[(k+k'-N)\tilde{g} -
\Omega -2n\chi\right]$,
\begin{equation}
G_{N}^n(t) = \sum_{k,k'=0}^N a_{k,k'} \sin (\tilde{\omega}_{k,k'}^n t)\,,
\label{eq:GN0}
\end{equation}
where
\begin{equation}
a_{k,k'} = 2^{-2N} \textstyle{\binom{N}{k} \binom{N}{k'}}
\left[(-1)^k-(-1)^{k'}\right]\,.
\end{equation}

Equation~(\ref{eq:BM_oscillations}) shows that $\langle B(t)\rangle$ is
characterized by many oscillations on timescales $\sim t_{\text{GHZ}}$,
since the $\tilde{\omega}_{k,k'}^n$ are of the same
order as $\Omega\gg\tilde{g},\chi$. However, the envelope indeed reaches its
maximum at $t_{\text{GHZ}}$, provided that only the ground state of the
resonator is significantly populated. These fast oscillations are the
manifestation of local rotations of the qubits,
Eqs.~(\ref{eq:rhotghz}-\ref{eq:localrotations}). We have seen that this
issue can be solved equivalently in two different ways and that the
state $|\text{GHZ}^+\rangle$ is indeed generated after
$t_{\text{GHZ}}$, either by applying some correcting pulse
$\exp(i\delta_N \hat J_z)$, defined in Eq.~(\ref{eq:correctingpulse}),
or by tuning the frequencies $\Omega$ and $\tilde{g}$ to satisfy the
condition Eq.~(\ref{eq:Omegaconditon2}). Assuming for simplicity
that $\delta_N=0$, we have then
\begin{equation}
G_{N}^{n}(t_{\text{GHZ}}) = \cos^{2N}\left(n\frac{\pi}{2}\frac{\chi}{\tilde{g}}\right) - \sin^{2N}\left(n\frac{\pi}{2}\frac{\chi}{\tilde{g}}\right).
\end{equation}
The fast oscillations of $\langle
B(t)\rangle$ around $t_{\text{GHZ}}$ become sharper as the number of
qubits $N$ increases, as shown in Fig.~\ref{fig:GNn}. In the simpler
case $\delta_N=0$, the behavior of $G_{N}^0$ around $t_{\text{GHZ}}$
is given by
\begin{equation}
G_{N}^0(t_{\text{GHZ}} + \tau) \simeq 1 - \tau^2 \frac{N
  \Omega^2}{4},\quad |\tau| \ll \frac{1}{\Omega}\,,
\end{equation}
and that also means that we need a higher precision, for larger $N$,
in controlling either the protocol time $t_{\text{GHZ}}$ or the
correcting pulse.

\begin{figure}
\centering
\includegraphics[width=\linewidth]{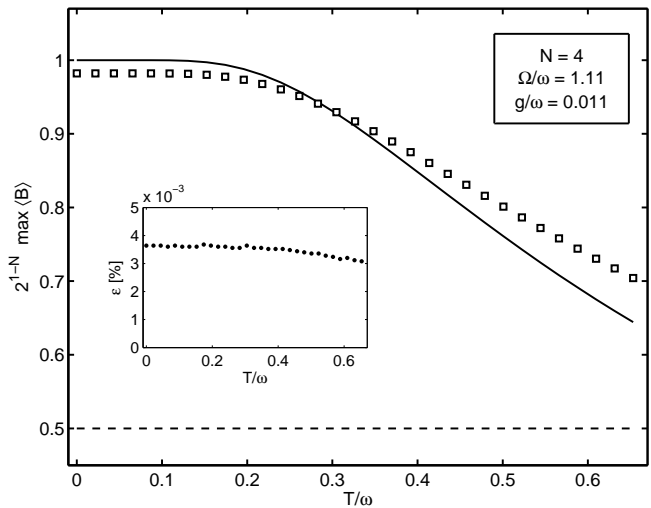}
\caption{Temperature dependence of the maximum $\max\langle B\rangle$
  of $\langle B(t)\rangle$, for $t\sim t_{\text{GHZ}}$ (squares),
  normalized by $2^{N-1}$. The solid line shows the theoretical bound
  $\tanh(\omega/(2T))$ for a resonator initially in the thermal state
  Eq.~(\ref{eq:rhothermal}).  Inset: relative deviation
  $\varepsilon=t_{\max}/t_{\text{GHZ}}-1$ of the time $t_{\max}$ at
  which $\max\langle B\rangle$ is realized compared to the predicted
  time $t_{\text{GHZ}} = \pi/(2\tilde g)$.  Here we considered $N=4$
  qubits and the parameters are $\Omega/\omega=1.105$,
  $g/\omega=0.0105$ and thus $g/\Delta\simeq 0.1$.  Local hidden-variable
  theory only allows values of $\langle B \rangle$ below the dashed
  line.  For $N=4$ this value also corresponds to the upper bound for
  biseparable states.}
\label{fig:Tdep_Detuning}
\end{figure}

Finally, the maximal value $\langle B(t_{\text{GHZ}})\rangle$ can reach
depends only on the initial state of the resonator
$\rho_{\text{res}}$, provided the above considerations have been taken
into account. Equations~(\ref{eq:correctedrho})
and (\ref{eq:definitionbm}) show that
\begin{equation}
\begin{aligned}
&2^{1-N} \text{Tr}\,\left[B ( e^{i\delta_N \hat J_z}
    \rho_{\text{qb}}(t_{\text{GHZ}}) e^{-i\delta_N \hat J_z} ) \right]
\\ & \;  =  \sum_{n=0}^{\infty} \langle
  n | \rho_{\text{res}}|n\rangle \left[ \cos^{2N}\left(n\frac{\pi}{2}\frac{\chi}{\tilde{g}}\right) - \sin^{2N}\left(n\frac{\pi}{2}\frac{\chi}{\tilde{g}}\right)\right]\,.
\end{aligned}
\end{equation}
For instance, we assume $\rho_{\text{res}}$ to be a thermal state
characterized by a temperature $T$,
\begin{equation}
\rho_{\text{res}} = \left(1-e^{-\omega/T}\right) \sum_n
e^{-n\omega/T} |n\rangle \langle n|\,.
\label{eq:rhothermal}
\end{equation}
In this simple case, the outcome of the Bell-Mermin operator measurement
$\langle B(t_{\text{GHZ}}) \rangle$ should be at least
$2^{N-1} \tanh(\omega/(2T))$.

A numerical evaluation of $\langle B(t)\rangle$, using the
Jaynes-Cummings Hamiltonian Eq.~(\ref{eq:JCham}), shows good
agreement with our theoretical analysis. We consider the ideal case of
homogeneous qubit and coupling frequencies and we choose frequencies satisfying
Eq.~(\ref{eq:Omegaconditon2}) such that $\delta_N=0$. We look for the
maximal value of $\langle B(t) \rangle$ around $t_{\text{GHZ}}$, that
is for $|t-t_{\text{GHZ}}|<\frac{\pi}{2\Omega}$, and for the time
$t_{\max}$ at which this maximal value is realized.
The results for $N=4$ qubits are shown in Fig.~\ref{fig:Tdep_Detuning}.

\section{Inhomogeneous coupling frequencies}
\label{inhomogeneous_coupling_frequencies}
To estimate whether our scheme is robust against small random
deviations in the physical parameters, we consider small
inhomogeneities in the coupling strengths $g_{j}^{(q)}$.  This
effect will be investigated numerically and, for this purpose we
compute the real-time evolution of the Bell-Mermin operator, using the
Jaynes-Cummings Hamiltonian Eq.~(\ref{eq:JCham}), truncated to the two
lowest levels of the transmon qubits.  This should capture the
main features of this effect, since in our effective description of the
system Eq.~(\ref{eq:effham0}), the third levels of the transmon qubits
only affect the ac-Stark shifts $\chi^{(q)}$ and renormalize the
resonator frequency. Assuming the qubit transition frequencies are
still homogeneous $\omega_{01}^{(q)} = \Omega$, the inhomogeneity of the
coupling frequencies $g_0^{(q)}$ produces inhomogeneous qubit-qubit
couplings coefficients 
$\tilde{g}_{00}^{(qq')} = |g_0^{(q)}g_0^{(q')}|/\Delta$.

The coupling constants $g_0^{(q)}$ are assumed to be Gaussian distributed
with mean $g$ and standard deviation $\delta g$. The notation $\{g_q\}$ denotes
a particular set of coupling frequencies $g_0^{(q)}$. The real-time evolution
of the Bell-Mermin operator for one set of coupling frequencies
$\{g_q\}$ is denoted $\langle B_{\{g_q\}}(t) \rangle$.

For a given number $n_r$ of random realizations $\{g_q\}$ ($n_r$
around 200) with fixed $\delta g$, we first calculate the mean value,
\begin{equation}
\langle \bar B(t) \rangle = \frac{1}{n_r} \sum_{\{g_q\}}\langle
B_{\{g_q\}}(t) \rangle\,.
\end{equation}
Then, the maximal value $\langle \bar B(t_{\max})\rangle$ defined by
\begin{equation}
\langle \bar B(t_{\max})\rangle = \max_{t\geq 0} \langle \bar
B(t)\rangle
\end{equation}
is found. Finally the variances, above and below the maximal mean value $\langle
\bar B(t_{\max})\rangle$, of the particular set $\left\{ \left\langle
B_{\{g_q\}}(t_{\max})\right \rangle \right\}$ are calculated. The variances are calculated
separately above and below, because the $\langle
B_{\{g_q\}}(t_{\max})\rangle$ are not Gaussian distributed. We
also calculate the median among the $\langle
B_{\{g_q\}}(t_{\max})\rangle$ and notice that the distribution is
strongly asymmetric.

\begin{figure}
\centering
\includegraphics[width=\linewidth]{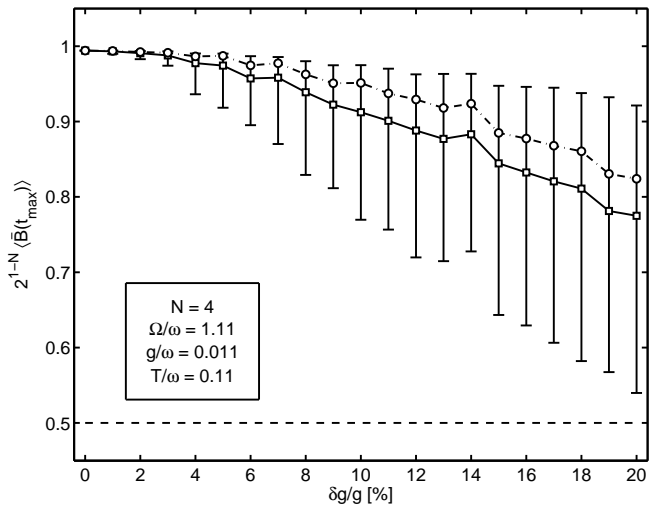}
\caption{Effect of inhomogeneous coupling frequencies $g_0^{(q)}$ with mean
  $g$ and standard deviation $\delta g$. We show the dependence of the
  maximal mean value $\langle \bar B(t_{\max})\rangle$ of $\langle
  B_{\{g_q\}}(t)\rangle$ on $\delta g/g$ for $t\sim t_{\text{GHZ}}$
  (squares).  The error bars show the standard deviation of $\langle
  B_{\{g_q\}}(t_{\max})\rangle$ above and below the mean value. The
  median of $\langle B_{\{g_q\}}(t_{\max})\rangle$ (circles) is
  clearly above the mean value.  Local hidden-variable theory only
  allows values of $\langle B \rangle$ below the dashed line. For
  $N=4$ this value also corresponds to the upper bound for biseparable
  states.}
\label{fig:inhomog}
\end{figure}

Results for $N=4$ and $\delta g/g$ between 0 to 20 \% are shown in
Fig.~\ref{fig:inhomog}. The time at which the maximum is attained is
generally in good agreement with the predicted value
$t_{\text{GHZ}}=\pi/(2\tilde{g})$, as long as $g/\Delta$ is small.
The value of $\langle\bar B(t_{\max})\rangle$ remains close
to the ideal one for $\delta g/g$ of the order of a few percents and
thus we notice that our scheme can tolerate some inhomogeneity in the
coupling constants.

\section{Conclusion}
To conclude, we have shown that it is possible to generate
multipartite Greenberger-Horne-Zeilinger states on a set of transmon
qubits in a circuit QED setup in a one-step deterministic protocol.
In the dispersive limit $g\ll\Delta$, such a system behaves as a fully
connected qubit network with exchange interactions proportional to
$\tilde{g}=g^2/\Delta$. The preparation time of the protocol is
inversely proportional to $\tilde{g}$.  The non-local nature of the
generated state can be investigated using a Bell-Mermin
inequality. Moreover, we have derived and applied bounds on the
expectation value of the Bell-Mermin operator as a detection criterion
for genuine $N$-partite entanglement.
Finally we have shown that our scheme is
robust against small inhomogeneities in the coupling frequencies. The
implementation proposed here looks like a promising way to generate
GHZ states, and hopefully can be experimentally realized in a circuit
QED setup.

\section{Acknowledgment}
We would like to thank S. Filipp and J. Koch for discussions and
correspondence about the qubit-qubit interaction of transmon qubits in
a circuit-QED setup, and for sending unpublished notes and M.
Pechal's Master thesis. We would also like to thank L. DiCarlo, S. Chesi,
F. Pedrocchi, and G. Str\"ubi for discussions. This work was
financially supported by the EC IST-FET project SOLID, the Swiss SNF,
the NCCR Nanoscience, and the NCCR Quantum Science and Technology.

\appendix*
\section{Schwinger representation of total spin operators}
\label{app:schwinger}

We present briefly the Schwinger representation
\cite{You2003PRL90} of the total spin operators $\hat
J_{x,y,z}=\frac{1}{2}\sum_q \sigma_{x,y,z}^{(q)}$. This turns out to be
particularly useful for calculations in the subspace of $\hat
J^2$-eigenstates with maximal eigenvalue
$\frac{N}{2}\left(\frac{N}{2}+1\right)$ where $N$ is the number of
spins. From now on we set $J\!=\!N/2$ and denote the states
$|J\!=\!N/2,J_{x,y,z}\rangle$ by $|J_{x,y,z}\rangle$.

States like $|J_z\rangle$ are sometimes referred to as Dicke states
\cite{Dicke1954PR93}, they form a complete basis of symmetric
$N$-qubit states, i.e., states invariant under any permutation of
qubits. We use for each qubit the standard basis
$\{|0\rangle,|1\rangle\}$ with the convention $\sigma_z^{(q)}|1\rangle_q =
|1\rangle_q$ and $\sigma_z^{(q)}|0\rangle_q =-|0\rangle_q$,
\begin{equation}
\begin{aligned}
&|J_z\!=\!k-N/2\rangle \\ & = \frac{1}{\sqrt{\textstyle{\binom{N}{k}}}}
  \sum_{p} |1\rangle_{p(1)} \cdots |1\rangle_{p(k)} |0\rangle_{p(k+1)}
  \cdots |0\rangle_{p(N)}\,,
\label{eq:dicke}
\end{aligned}
\end{equation}
with $0\leq k \leq N$ and where the sum is taken over the $\binom{N}{k}
 = \frac{N!}{k!(N-k)!}$ nonequivalent possible permutations $p$
that give different product states.

The operators $\hat J_i$ are defined by means of two independent bosonic
operators $a$ and $b$, with commutation relations $[a,a^{\dag}]=[b,b^{\dag}]=1$
and $[a,b]=[a,b^{\dag}]=0$,
\begin{equation}
\begin{gathered}
\hat J_x = \frac{1}{2}(b^{\dag}a + a^{\dag}b)\,,\\
\hat J_y = \frac{1}{2i}(b^{\dag}a - a^{\dag}b)\,,\\
\hat J_z = \frac{1}{2}(b^{\dag}b - a^{\dag}a)\,,
\end{gathered}
\end{equation}
fulfilling the SU(2) algebra $[\hat J_l,\hat J_m]=i\epsilon_{lmn}\hat
J_n$. Eigenstates of $\hat J_z$ can be expressed as
\begin{equation}
 |J,J_z\rangle = \frac{\left(b^{\dag}\right)^{J+J_z}
 |\left(a^{\dag}\right)^{J-J_z}}{\sqrt{(J+J_z)!(J-J_z)!}}
 |n_a\!=\!0,n_b\!=\!0\rangle\,,
\end{equation}
where $|n_a\!=\!0,n_b\!=\!0\rangle$ is the vacuum state of the
operators $a$ and $b$. Since the choice of the operators $a$ and $b$ is not
unique, we can equivalently introduce the operators $c= (a -
b)/\sqrt{2}$ and $d= (a + b)/\sqrt{2}$, leading to $\hat J_x =
\frac{1}{2}( d^{\dag} d - c^{\dag} c )$ and
\begin{equation}
|J,J_x\rangle = \frac{\left(d^{\dag}\right)^{J+J_x}
|\left(c^{\dag}\right)^{J-J_x}}{\sqrt{(J+J_x)!(J-J_x)!}}
|n_a\!=\!0,n_b\!=\!0\rangle\,.
\end{equation}
We straightforwardly obtain the decomposition of the states
$|J,J_x\rangle$ in terms of $|J,J_z\rangle$ and in particular
\begin{align}
|J_x\!=\!\pm N/2\rangle & = \bigotimes_{q=1}^N \frac{|0\rangle_q \pm
|1\rangle_q}{\sqrt{2}} \nonumber
\\ &= \frac{\left( a^{\dag} \pm b^{\dag}\right) ^N}{\sqrt{ 2^N N!}}
|n_a\!=\!0,n_b\!=\!0\rangle \\ &= \frac{1}{2^{N/2}} \sum_{k=0}^N (\pm
1)^{k} \sqrt{\textstyle{\binom{N}{k}}}
|J_z\!=\!k\!-\!N/2\rangle\,. \nonumber
\end{align}

Defining the ladder operators $\hat J_{\pm} = \hat J_x \pm i \hat J_y$
of the total spins, the Dicke states can also be expressed as
\begin{equation}
\begin{aligned}
|J_z\!=\!k\!-\!N/2\rangle & = \frac{\left(\hat J_+\right)^k}
{k!\sqrt{\textstyle{\binom{N}{k}}}} \bigotimes_{q=1}^N |0\rangle_q
\\ & = \frac{\left(\hat J_-\right)^{N-k}}{(N-k)! \sqrt{\textstyle{\binom{N}{k}}}}
 \bigotimes_{q=1}^N |1\rangle_q\,.
\end{aligned}
\end{equation}

\phantom{-- the end --}


\begin{thebibliography}{35}
\bibitem{note_koch} This can be seen by applying perturbation theory
  in $\tilde g_{jj'}^{(qq')}$ for $|j-j'|>1$ to Eq.~(\ref{eq:canonical}). 
We would like to thank Jens Koch for pointing out this argument.
\end{thebibliography}
\end{document}